\documentclass[aps,prl,twocolumn,groupedaddress]{revtex4} 

\usepackage{graphicx}
 
\begin{document} 
 
\title{New Pseudo-Phase Structure for $\alpha$-Pu} 
 
 
 
\author{J. Bouchet$^1$, R. C. Albers$^1$, M. D. Jones$^2$ and G. Jomard$^3$} 
\affiliation{$^1$Los Alamos National Laboratory, Los Alamos, NM 87545\\
$^2$Department of Physics and Center for Computational
Research,
University at Buffalo, The State University of New York,
Buffalo, NY 14260\\
$^3$Centre d'etudes de Bruyeres le Chatel, France.} 
 
 
\date{\today} 
 
\begin{abstract} 
 
In this paper we propose a new pseudo-phase crystal structure,
based on an orthorhombic distortion of the diamond structure,
for the ground-state $\alpha$-phase of plutonium.  Electronic-structure 
calculations in the generalized-gradient approximation give approximately
the same total energy for the two structures. Interestingly, our new pseudo-phase
structure is the same as the Pu $\gamma$-phase 
structure except with very different b/a and c/a ratios. We show how the
contraction relative to the $\gamma$ phase, principally in the $z$ direction,
leads to an $\alpha$-like structure in the [0,1,1] plane. This is an important 
link between two complex structures of plutonium and opens new 
possibilities for exploring the very rich phase diagram of Pu through 
theoretical calculations. 
 
\end{abstract} 
 
 
 
 
 
\maketitle  
 
Pure plutonium (which also includes Pu with small amounts of other 
solute atoms) is an exotic metal that is poorly understood.  It 
appears to be going through a Mott-insulator transition in its $5f$ 
electronic orbitals as it transforms into different crystal structures 
as a function of temperature.  Actinide elements lighter than Pu have 
itinerant $5f$ electrons; elements heavier than Pu form a second 
rare-earth series with localized $5f$ electrons.  The $\alpha$-Pu phase 
(zero-temperature ground state) is believed to have mainly itinerant 
$5f$ electrons.  However, $\delta$-phase (fcc) Pu is believed to be in 
neither limit (neither completely itinerant nor localized), but is 
instead in some kind of exotic electronic state that we may describe 
as partially localized, for want of any better terminology to apply to 
this system\cite{Albers01}. 

The understanding of this phase transformation, with its associated large 
volume per atom expansion, has been an exciting topic of discussion 
within band-structure theory and related correlated-electron 
extensions to it for several decades now. But almost all of the 
more sophisticated recent calculations have been done 
for the fcc $\delta$ phase\cite{Savrasov00,Savrasov01,bouchet,Soderlind02}.   
This crystal structure has one atom per unit cell and is amenable to better,
more computationally intensive techniques for handling the 
correlated electron physics, which is believed to be the fundamental 
issue controlling the $f$-shell Mott-insulator transformation. 
Because the physics is so difficult, we believe that studying the 
strongly correlated $\delta$ phase in isolation is less useful than also
including the more weakly correlated 
$\alpha$ phase in the study.  A correct theory should get both limits 
right, and is a strong constraint on the theoretical approaches that 
are used.  The chief problem with including the $\alpha$ structure is 
its complicated crystal structure, which is a low symmetry monoclinic
structure with space group \textit{P21/m} and involves 16 
atoms per unit cell\cite{Zachariasen63,Espinosa01}.
Because of this complexity, for example, a fully 
relaxed structure (volume and internal parameters) has not yet been 
obtained with DFT. 

We believe that we have now found a solution to this problem.  In this 
paper we propose a new pseudo structure for $\alpha$ Pu that involves 
only two atoms per unit cell and is a relatively symmetrical orthorhombic
crystal structure. 
We show that the real $\alpha$ crystal structure only involves small 
distortions of various atoms around this pseudo structure and that the 
electronic-structure energy of both systems (at least at the 
band-structure level) is nearly identical. 

By a pseudo structure we mean a simpler crystal structure where the 
atoms are quite close to the positions of the more complex real 
structure.  Pseudo-structures are interesting for two reasons: (1) 
they provide insight into the geometrical structure of the atoms in 
the original structure and possible phase-transformation pathways, and 
(2) they provide an alternative less expensive crystal structure for 
complicated electronic and microstructural calculations.  From a 
physics point of view, one can view the pseudo structure as an 
approximation to the real structure that captures the dominant total 
energy of the electronic-structure (heat of formation) and local 
bonding effects; the real structure can then be viewed as just a minor 
distortion of the pseudo structure, with only a small energy 
difference between the two structures. 

In the case of Pu, insight into the crystal structure is extremely 
important.  Unlike, for example, the cubic diamond structure of Si 
where it is easy to understand the tetrahedral nearest-neighbor 
structure as a consequence of directional s-p bonds, or the close 
packed fcc (face-centered-cubic) crystal structures favored by 
Madelung energies in metals, it is extremely difficult to understand 
why the lowest temperature Pu phases are so extremely anisotropic and 
complex.  In $\alpha$ Pu, for example, the nearest-neighbor structure 
involves a complicated arrangement of short and long bonds\cite{Donohue}. 
 How can we understand this, and why do the crystal structures 
become more isotropic and metallic like at high temperatures?  Good 
pseudo structures give us the opportunity to look at these crystal 
structures in new ways that may eventually stimulate insights that 
will resolve some of the questions. 

Understanding the complex phase diagram of Pu is likely 
to involve consideration of strong directional bonding and high 
density of states effects of the $f$ electrons and how that bonding 
weakens under strong electron-electron correlations.  Sophisticated 
electronic-structure calculations are necessary to make progress in 
this area.  Unfortunately, the $\alpha$ and $\beta$ phases, with 16 and 34 
atoms in the unit cell, respectively, are just too computationally 
expensive for such calculations to be practical. To have a simple
pseudo-phase crystal structure 
that can mimic the actual bonding arrangement of the extremely 
complicated crystal structure of the real phase makes it possible to 
do these very sophisticated electronic-structure calculations with 
little loss of accuracy or insight. If accurate pseudo phases exist, 
they also make possible complicated 
microstructural models of the phase transformations that captures the 
dominant components of free energy differences without requiring huge 
numbers of parameters. For example, Ginzburg-Landau
models become 
much more accessible through the use of pseudo phases.  An attempt is 
currently being made to do this for the very important $\alpha$ to 
$\delta$ phase transformation of Pu\cite{Saxena}. 

What is even more interesting is that our proposed pseudo structure is 
identical with the $\gamma$ phase of Pu, except that the unit cell 
dimensions are greatly changed.  This crystal structure is an 
orthorhombic distortion of the cubic diamond lattice.  It has two 
atoms in the smallest unit cell, one at the origin, and the second 1/4 
of the way along the a, b, and c axes (which are now all different in 
the orthorhomic unit cell).  Because both atoms have a face-centered 
configuration, there are 8 atoms in the primitive orthorhombic unit 
cell (just like cubic diamond).  As we will show, our pseudo 
structure is far better than the only other previously proposed pseudo 
structure for $\alpha$ Pu, the Crocker pseudo structure\cite{Crocker71}.
This structure was based on the insight that the 
the $\alpha$ structure can be viewed as two repeating planes of a 
distorted hexagonal structure.  Although this structure also has
two atoms in its unit cell (each of the distorted hexagons are replaced
by symmetric hexagons), it is also monoclinic, and has not proved
useful for electronic structure calculations.  We will also show that
its near-neighbor shells are a poor approximation to the real
$\alpha$ phase. 

What makes Pu fascinating from an electronic-structure point of view 
is the instability in the $5f$ shell, which have itinerant electrons 
for $\alpha$ Pu, and localize into the core for Am and higher-Z 
actinides.  This Mott-like transition for the $5f$ electrons involves 
very strong electron-electron correlations, and requires a much more 
complicated theory for the electrons than the usual local density 
approximation (LDA) for density functional theories (DFT), which are 
like a mean-field treatment of the correlations\cite{Lundqvist83}. 
If the parabolic decrease of the volume of the light actinides as a 
function of atomic number can be explained with a simple Friedel model, 
plutonium deviates from this theoretical estimation, since Pu 
has a larger volume than Np. Until now there is no definitive 
explanation of this upturn.  In addition, it is well known that DFT in 
its local density or generalized gradient approximation (GGA), 
due to the presence of strong correlation effects, has been unable to 
describe the $\delta$-phase of plutonium, underestimating its atomic 
volume by about 20\%, for example. But recently, several attemps to go 
beyond LDA, have given some understanding of the $\alpha$-$\delta$ 
transition. The LDA+$U$ method\cite{bouchet,Savrasov00} and the 
LDA+DMFT (dynamical mean-field theory)\cite{Savrasov01}, by taking into 
account the localization of the $f$-states, have shown the role of 
spin and orbital polarization in describing the $\delta$-phase. 

S\"oderlind \textit{et al}\cite{Soderlind97} 
have performed \textit{ab initio} relativistic total energy 
calculations and offered an explanation for the crystal structure of $\alpha$-Pu in 
terms of a Peierls-Jahn-Teller-like distortion. Comparing the 
total energy of the $\alpha$ phase with total energies of symmetric 
(fcc, bcc, hcp) structures and other actinides-like structures 
(such as $\alpha$-U and $\alpha$-Np) they showed that the $\alpha$-Pu 
structure minimizes the total energy for plutonium, but they couldn't 
explain the upturn in equilibrium volume for Pu. More recently Jones 
\textit{et al}\cite{Jones00} have performed calculations with two different 
methods, FLAPW and LCGTO-FF for all the light actinides. Although their 
theoretical volumes agree well (within about 2\%) with the experimental 
volumes for Th to Np, they show a larger discrepancy (7\%) for 
$\alpha$-Pu, and no upturn between Np and Pu.  No attempt was made to 
relax the structural parameters of the $\alpha$ phase. 
They also speculated 
that some localization effects could already be present in $\alpha$-Pu 
accounting for the discrepancy between the computed and measured 
atomic volume and causing the upturn.  
To our knowledge, no fully relaxed electronic 
structure calculations have been performed on the $\alpha$, $\beta$ and the 
$\gamma$ phase and a complete and comprehensible picture of the 
plutonium phase diagram still does not exist. 

In our calculations, we have mainly used the projector augmented-wave
(PAW) method, which combines ideas from 
pseudopotentials and the linearized augmented-plane-wave (LAPW) method
\cite{wien}. 
Forces and the full stress tensor can be easily calculated and used to 
relax atoms into their instantaneous ground state.  This is implemented in
the plane-wave basis VASP code\cite{Kresse96} 
with spin-polarized DFT and the GGA of Perdew and Wang\cite{Perdew92}.
Ultrasoft Vanderbilt pseudopotentials were used to represent the core 
electrons. 

This allowed us to perform for the first time a fully relaxed calculation of the 
$\alpha$-phase.  Results are presented in Table~\ref{param}. As in 
previous calculations\cite{Soderlind97, Jones00}, the atomic volume is 
slightly lower compared to the experimental value. This discrepancy is 
partially due to the absence of spin-orbit coupling in our 
calculations\cite{Kollar97,Jones00}. The relaxed structure is 
very closed to the experimental one (the energy difference is about 4 mRy),  
and the positions of the atoms 
relaxed by only 2 or 3 \%, showing that DFT-GGA reproduces the ground 
state of plutonium with a good precision.  

We have compared the total energies of the $\alpha$ structure, 
Crocker's structure, and the $\gamma$ structure. As expected, the $\alpha$ phase
has the lowest energy, while Crocker's structure is about 20 mRy larger, and
the $\gamma$ phase 36 mRy larger. 
The theoretical volume of the $\gamma$ phase is very 
far from the experimental volume (23.14 \AA$^3$), but very close to the experimental 
volume of the $\alpha$ phase, see Table~\ref{param}. 
If we minimize the total energy of this structure as a 
function of $b/a$ and $c/a$ we obtain a large contraction in the $z$ 
direction and a smaller one in the $y$ direction to obtain a structure 
very different from $\gamma$ (we will call this new structure 
``pseudo-$\alpha$" in the rest of the paper). The volume increases slightly and
is roughly 
equal to the theoretical volume for the $\alpha$ 
structure, see Table~\ref{param}. Surprisingly the total energy is also very close to the total 
energy obtained for the $\alpha$ structure, about 3 mRy higher, almost within the accuracy of calculations. Using a full-potential linear  
augmented-plane wave method (FLAPW) we have checked the influence
of the spin-orbit coupling on our results (see Table~\ref{param}).
The total energy of the pseudo-structure is still very close to the $\alpha$ one and 
the spin-orbit coupling doesn't help to differentiate the two structures.

It is interesting to speculate on why we find nearly the same 
total energy for two structures that appear to be so completely
different. To understand this it is useful to compare the interatomic distances in 
the $\gamma$, the pseudo-$\alpha$, and the $\alpha$ structure,
which are shown in Table~\ref{nn}. Although the atomic environment around each 
atom in the $\alpha$ structure is very complex, we can divided it into roughly 
two categories: short and long bonds, well separated.  Each atom has four first-nearest 
neighbors and ten second-nearest neighbors, except atom I which 
has 3 and 13 and atom VIII which has 5 and 7. During the transformation
from $\gamma$ to pseudo-$\alpha$ the first 
shell of neighbors contracts around the atom while the second and the 
third shell expand to be very close to the fourth shell. This gives a 
local environment of 4 nearest neighbors and 10 second-nearest neighbors
for the pseudo-$\alpha$ , 
exactly what is found in average in the $\alpha$ structure. In addition, the interatomic  
distances are very close. So the relaxation of the $\gamma$ phase gives us a new structure,  
with nearly exactly the same distribution of interatomic distances than in the $\alpha$ phase.  
Crocker's structure, however,  is quite different with only  
two first-nearest neighbors, and an intermediate shell with four atoms. So, the
local atomic environment is more important than the monoclinic symmetry
in determining the energy of the $\alpha$ structure. 

\begin{figure} 
 
\rotatebox{270}{\includegraphics[scale=0.4]{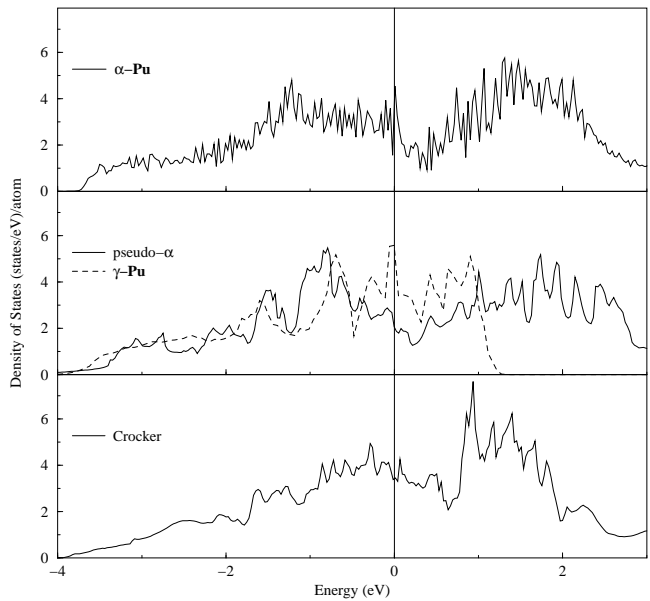}} 
 
\caption{\label{dos}Density of states for $\alpha$-Pu (upper pannel), 
pseudo-$\alpha$ and $\gamma$-Pu in dashed (middle panel), and 
Crocker's structure (lower panel) calculated at the theroretical 
equilibrium volume.} 
 
\end{figure}

In Fig.~\ref{dos}, we show the density of states (DOS) obtained for the 
three structures at their theoretical volumes. They are very similar, 
with the Fermi level in a pseudo gap and two broad features, one above 
and one below the Fermi level. The DOS of the $\gamma$-Pu with the 
experimental parameters has been compared to the pseudo-$\alpha$ structure and 
we can see the presence of a peak at the Fermi energy, so during the 
contraction of the cell along the $b$ and $c$ axis there is the 
opening of a large pseudo gap with the reduction of the density of 
states at the Fermi level.  

\begin{figure} 
 
\rotatebox{0}{\includegraphics[scale=0.4]{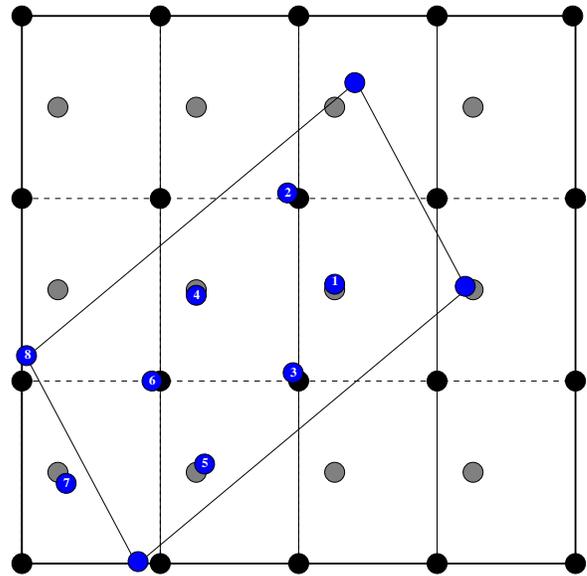}} 
 
\caption{\label{gamal}Distorted diamond structure (pseudo-$\alpha$) in the (011) 
direction. The blue atoms are the eight positions in the $\alpha$ 
structure.} 
 
\end{figure}

Looking at the [011] plane of the pseudo-$\alpha$ structure, see
Fig.~\ref{gamal}, one can almost visualize the $\alpha$ phase.  The atoms in dark  
and grey are the two distinct atoms in the pseudo-$\alpha$ structure,
and we have superimposed in blue the $\alpha$-structure atoms. 
They are very similar despite the complexity of 
the $\alpha$ pattern and the relative simplicity of the distorted diamond 
structure. We have calculated the distance between related atoms
in the two different crystal structures, which has an average of about 0.25~\AA.
These distortions are responsible for the small total-energy difference  
between the two structures.    
In addition, the [011] planes have the same interlayer distance as  
in the $\alpha$ phase (2.31~\AA\ and 2.26~\AA\ for pseudo-$\alpha$ and $\alpha$ respectively). 
As we indicated earlier the new structure respects the average 
surrounding of an atom in the $\alpha$ phase, with four nearest 
neighbours, but exactly as in the $\alpha$ phase, there is two in the same plane,
one in the plane above and one in 
the plane below. The angle between one atom and its nearest neighbours from the two other planes is also very similar  
(around 137$^\circ$ and $130^\circ$ for pseudo-$\alpha$ and $\alpha$
respectively, in diamond it is 109$^\circ$). 
 In addition, the connection of our pseudo 
phase with the gamma phase suggests a deep structural relationship 
between the low-temperature phases of Pu that needs further 
examination.  It is interesting to observe that the b/a and c/a ratios 
of the real $\gamma$ phase of Pu have a tendency to move toward 
the values for our new pseudo phase as the temperature is 
lowered.  Zachariasen and Ellinger measured the lattice constants in function of temperature. 
The lengths of $b$ and $c$ both decrease with temperature, while the length of $a$ increases\cite{Donohue}. 
If the $\beta$ phase did not interrupt this trend, it is 
likely that $\gamma$ Pu would slowly continue in this way until it 
would reach the phase transformation into the $\alpha$ phase that would 
then require just a tiny distortion.  

In conclusion we have discovered a very simple pseudo structure reproducing the
main features of the complex ground state of plutonium.
Our results show that the local atomic environment  
in the $\alpha$ structure with four first-nearest neighbors and 10 second-nearest
neighbors is its crucial  
characteristic. We have found a very simple path (following the experimental
variation of the lattice constants of $\gamma$ Pu with temperature)
between  $\gamma$ and $\alpha$ plutonium.
These results show the predictive power of DFT calculations and our new 
pseudo structure of the $\alpha$ phase opens new possibilities for 
exploring the behavior of plutonium.  It will allow approximate calculations
of the $\alpha$ phase with far more sophisticated techniques than have been previously possible. 
 
\begin{table}
 
 
\caption{\label{param}Structural and internal parameters (\AA) for 
the experimental and theoretical $\alpha$, Crocker, pseudo-$\alpha$ and 
$\gamma$ structures of plutonium. For $\alpha$, the 
volume and all the internal parameters have been relaxed. The energies (mRy) are relative to the  
$\alpha$ structure, which is the ground state for all the calculations,  
with or without spin-orbit coupling.} 
\begin{ruledtabular} 
 
\begin{tabular}{cccccc} 
 
&\multicolumn{2}{c}{Pu-$\alpha$}&Crocker&$\gamma$&pseudo-$\alpha$\\ 
 
\hline 
 
&Exp.&Th.&&\\ 
 
\hline 
 
a&6.19&5.96&2.94&2.85&3.50\\ 
 
b/a&0.78&0.76&1.56&1.83&1.61\\ 
 
c/a&1.77&1.83&1.93&3.22&2.09\\ 
 
$\beta$&101.8&101.8&116.3&\\ 
Volume&20.0&18.0&17.3&17.0&18.0\\ 
Energy (GGA)&&0&20&36&3\\ 
Energy (GGA+SO)&&0&18&&3 
\end{tabular} 
 
\end{ruledtabular} 
 
\end{table} 
\begin{table}
 
\caption{\label{nn}Nearest neighbours distances (\AA) for the 
Crocker structure, the $\gamma$-Pu, the pseudo-$\alpha$, and the 
$\alpha$-Pu. These values are given for the theoretical equilibrium 
volume for each structure.} 
 
\begin{ruledtabular} 
 
\begin{tabular}{cccccc} 
 
&\multicolumn{5}{c}{distances (number)}\\ 
 
\hline 
 
&Crocker&$\gamma$-Pu&pseudo&$\alpha$&\\ 
 
\hline 
 
1 n.n&2.44(2)&2.73(4)&2.47(4)&2.46-2.55(4)& short bonds\\ 
\cline{2-6}
 
2 n.n&2.79(2)&2.85(2)&3.31(4)&&\\ 
 
3 n.n&2.79(2)&2.97(4)&3.50(2)&3.21-3.59(10)&long bonds\\ 
 
4 n.n&3.02(2)&3.39(4)&3.49(4)&&\\ 
 
\end{tabular} 
 
\end{ruledtabular} 
 
\end{table}

\bibliography{biblio} 
 
\end{document}